\documentclass[sigconf,authorversion,nonacm]{acmart}

\usepackage{subfig}
\usepackage{textgreek}
\usepackage{dirtytalk}
\usepackage{enumitem}
\usepackage{amsthm}
\usepackage{hyperref}

\newtheoremstyle{boldhyp} % name
    {0em}                    % Space above
    {0em}                    % Space below
    {\itshape}                    % Body font
    {\parindent}                  % Indent amount
    {\bfseries}           % Theorem head font
    {.}                   % Punctuation after theorem head
    {.3em}                % Space after theorem head
    {}                    % Theorem head spec (can be left empty, meaning ‘normal’)

\theoremstyle{boldhyp}
\newtheorem{H1}{H}

\newlist{questions}{enumerate}{2}
\setlist[questions,1]{label=\textbf{RQ:},ref=\textbf{RQ}}

\copyrightyear{2024} 
\acmYear{2024} 
\setcopyright{acmlicensed}
\acmConference[MuC '24]{Proceedings of Mensch und Computer 2024}{September 1--4, 2024}{Karlsruhe, Germany}
\acmBooktitle{Proceedings of Mensch und Computer 2024 (MuC '24), September 1--4, 2024, Karlsruhe, Germany}
\acmDOI{10.1145/3670653.3670680}
\acmISBN{979-8-4007-0998-2/24/09}

\begin{document}

%%
%% TITLE
%%\title{What did I say again? Relating User Needs to Product Attributes in Conversational Commerce}

%% \title{What did I say again? Relating User Needs to Search Outcomes in Product Search}

\title{What Did I Say Again? Relating User Needs to Search Outcomes in Conversational Commerce}

%% \title{What did I say again? Relating User Needs to Product Attributes in Conversational eCommerce}

%%
%% AUTOREN
\author{Kevin Schott}
\email{kevin.schott@gesis.org}
\affiliation{%
 \institution{GESIS – Leibniz Institute for the Social Sciences}
 \city{Cologne}
 \country{Germany}
 }

\author{Andrea Papenmeier}
\email{a.papenmeier@utwente.nl}
\affiliation{%
 \institution{University of Twente}
 \city{Enschede}
 \country{Netherlands}
 }

\author{Daniel Hienert}
\email{daniel.hienert@gesis.org}
\affiliation{%
 \institution{GESIS – Leibniz Institute for the Social Sciences}
 \city{Cologne}
 \country{Germany}
 }

\author{Dagmar Kern}
\email{dagmar.kern@gesis.org}
\affiliation{%
 \institution{GESIS – Leibniz Institute for the Social Sciences}
 \city{Cologne}
 \country{Germany}
 }

%%
%% SHORT AUTHORS
\renewcommand{\shortauthors}{Schott et al.}
 
%%
%% ABSTRACT

\begin{abstract}
Recent advances in natural language processing and deep learning have accelerated the development of digital assistants. In conversational commerce, these assistants help customers find suitable products in online shops through natural language conversations. During the dialogue, the assistant identifies the customer's needs and preferences and subsequently suggests potentially relevant products. Traditional online shops often allow users to filter search results based on their preferences using facets. Selected facets can also serve as a reminder of how the product base was filtered. In conversational commerce, however, the absence of facets and the use of advanced natural language processing techniques can leave customers uncertain about how their input was processed by the system. This can hinder transparency and trust, which are critical factors influencing customers' purchase intentions. To address this issue, we propose a novel text-based digital assistant that, in the product assessment step, explains how specific product aspects relate to the user's previous utterances to enhance transparency and facilitate informed decision-making. We conducted a user study (N=135) and found a significant increase in user-perceived transparency when natural language explanations and highlighted text passages were provided, demonstrating their potential to extend system transparency to the product assessment step in conversational commerce.
\end{abstract}

%%
%% CCSXML


\begin{comment}
\begin{CCSXML}
<ccs2012>
   <concept>
       <concept_id>10003120.10003121</concept_id>
       <concept_desc>Human-centered computing~Human computer interaction (HCI)</concept_desc>
       <concept_significance>300</concept_significance>
       </concept>
   <concept>
       <concept_id>10002951.10003317.10003325.10003327</concept_id>
       <concept_desc>Information systems~Query intent</concept_desc>
       <concept_significance>500</concept_significance>
       </concept>
   <concept>
       <concept_id>10003120.10003121.10003122.10003334</concept_id>
       <concept_desc>Human-centered computing~User studies</concept_desc>
       <concept_significance>500</concept_significance>
       </concept>
   <concept>
       <concept_id>10003120.10003121.10003124.10010865</concept_id>
       <concept_desc>Human-centered computing~Graphical user interfaces</concept_desc>
       <concept_significance>300</concept_significance>
       </concept>
   <concept>
       <concept_id>10003120.10003145.10011769</concept_id>
       <concept_desc>Human-centered computing~Empirical studies in visualization</concept_desc>
       <concept_significance>500</concept_significance>
       </concept>
 </ccs2012>
\end{CCSXML}

\ccsdesc[300]{Human-centered computing~Human computer interaction (HCI)}
\ccsdesc[500]{Information systems~Query intent}
\ccsdesc[500]{Human-centered computing~User studies}
\ccsdesc[300]{Human-centered computing~Graphical user interfaces}
\ccsdesc[500]{Human-centered computing~Empirical studies in visualization}

\end{comment}

%%
%% KEYWORDS

%%
%% DISPLAY HEADER
\maketitle

\title{Relating User Needs to Search Outcomes in Conversational Commerce}

%% ============================================================================= MAIN BODY

\section{Introduction}
\label{intro}
In recent years, we have witnessed rapid advances in natural language processing (NLP) and deep learning techniques, resulting in the rise of powerful generative Artificial Intelligence (AI) such as GPT-4~\cite{Achiam2023}. This has led to the proliferation of conversational user interfaces (CUIs) such as virtual (voice) assistants and chatbots, which allow users to interact with computers in natural human language (NL)~\cite{AccessibleCUIS, McTear}. The growing popularity of CUIs can also be observed in the context of online shopping. \textit{Conversational commerce}, a term which was introduced in 2015~\cite{Messina, Tuzovic2018} and defined by Balakrishnan et al. as ``buying activity by a customer through a digital assistant''~\cite{ConversationalCE} has emerged as a new option for companies to sell products to their customers. These digital (shopping) assistants engage customers in a voice or chat dialogue to elicit their needs and preferences and find a suitable product. The conversation mimics a natural dialogue with a human shop assistant intended to gather necessary information and to build trust~\cite{Tsagkias2021}.

% detailed problem
The graphical user interfaces of traditional online shops often display facets that allow customers to specify their preferences and filter the search results accordingly. Additionally, selected facets can remind users of the applied filters. In conversational commerce, however, facets are missing, and digital assistants usually employ NLP techniques to interpret the users' utterances. This can lead to users questioning how the system has processed their input to come up with a recommended product~\cite{Pu_2012, Herlocker}.

% therefore we did...
To address this issue, we investigate how explanations that indicate how user utterances were mapped to product attributes can support users during product assessment in conversational commerce. We understand explanations to mean ``[a]ny feature or aspect that enhances the interpretability and transparency of the system, making it more understandable to [users]''~\cite{Juneja2024}. We introduce a text-based digital assistant for laptop search that maps vague user responses regarding different laptop attributes to specific values. Inspired by faceted search, the assistant explains this mapping in the result presentation. We conducted a user study and analyzed participants' perceptions of two types of result explanations (namely text snippet highlighting and NL explanations) compared to a baseline condition without any explanations. The study's findings indicate that providing NL explanations together with text snippet highlighting for how user responses were mapped to different attributes of a recommended product can enhance users' perceived transparency and are deemed helpful by users.

\section{Related Work}
\label{relatedWork}
As the interaction with digital assistants involves aspects of both search and recommendation \cite{Yongfeng2018}, we discuss related literature from the research fields of information retrieval (IR) and recommender systems (RSs) in addition to the limited research on explanations specific to conversational commerce. Thus, users who search for a suitable product by conversing with a digital assistant receive either a personalized product suggestion or a list of search results. In the following, we have a closer look at the goals of explanations, their effects on perceived transparency and decision-making, as well as text highlighting and NL explanations as explanatory components.

\subsection{Goals of Explanations}
\label{subsec:goals}
 Regarding the goals of explanations in RSs, Tintarev~\cite{Tintarev2007} describes seven possible aspects: Transparency (explaining how the system works), scrutability (allowing users to tell the system that it is wrong), trustworthiness (increasing users' confidence in the system), effectiveness (helping users make decisions), persuasiveness (convincing users to try or buy), efficiency (helping users to make decisions faster), and satisfaction (improving the ease of use or enjoyment). Focusing on the first aspect, Vorm and Miller~\cite{Vorm2018} propose the five-factor model of transparency in RSs, which includes the factors Qualities of Data (e.g., ``What are the sources of data?''), Options (e.g., ``Choices known to the system are made available to the user''), User Representation (e.g., ``Does the system know and consider the user in its model?''), Social Influence (e.g., ``How is the user grouped with others?''), and System Parameters and Logic (e.g., ``Information about system logic, reasoning, policies, limitations, etc.''). Tintarev and Masthoff~\cite{Tintarev2015} note that the terms ``explanation'' and ``justification'' are often used interchangeably. However, Vig et al.~\cite{Vig} take a more detailed approach. They distinguish between (result) justification, which explains why a particular item is recommended (e.g., ``This movie is a good choice because you liked other movies by this director''), and (system) transparency, which clarifies how the recommendation mechanism operates (e.g., ``My algorithm analyzes your viewing history and identifies patterns, such as your preference for this director’s movies''). In this work, we focus on system transparency.

\subsection{Effects of Explanations on Perceived Transparency and Decision-making}
\label{subsec:effects}
Tsagkias et al.~\cite{Tsagkias2021} point out that explanations in conversational commerce can increase transparency by enabling users to understand what data from their input is being processed and how the search or recommendation mechanism works. Jin et al. \cite{Jin2024} developed the digital shopping assistant ``PhoneBot'' for mobile phones that provides users with NL explanations that justify product recommendations within the chat. Based on the users' expressed preferences, the assistant explains how well a suggested mobile phone ranks within the online shop's product library. The study shows that these explanations can enhance users' perceived transparency by improving explainability. In a user study with a digital assistant for hotel search, Hernandez-Bocanegra and Ziegler \cite{Hernandez2023} found that enabling users to interact with explanations by requesting individual customer comments can increase their perception of transparency through enhanced perception of explanation quality compared to the sole provision of aggregated customer opinions.

In RSs, Pu et al.~\cite{Framework} show that explanations can increase users' perception of transparency, enhancing users' trust and confidence in decision-making. Cramer et al.~\cite{Cramer2008} demonstrate that allowing users to view the criteria based on which an art recommender makes its suggestions can enhance users' perceived and actual understanding of how the system works. In their user study, Gedikli et al.~\cite{Gedikli} compared different explanation styles for a movie recommender. They discovered that personalized tag clouds color-coded to represent users' sentiment towards tags such as ``politics'', ``drama'', or ``classic'' can support users in decision-making, ultimately improving recommendation effectiveness and increasing users' perceived level of system transparency. In another study for a movie recommender, Vig et al.~\cite{Vig} provided users with community tags for a recommended movie and predicted their perceived relevance and preference for those tags. They show that this type of explanation can improve users' perceived justification for ratings predicted by the RS and support their decision-making. In the context of music recommendation, Millecamp et al.~\cite{Millecamp} found that explanations could enhance users' perception of the recommendations' effectiveness and improve users' understanding. Furthermore, a user study by Verbert et al.~\cite{Verbert} demonstrates that visually explaining and allowing users to explore publication recommendations in the context of academic conferences can increase their effectiveness. Investigating personalized NL explanations in a news RS, ter Hoeve et al. found that they may increase users' trust~\cite{TerHoeve2017}.

In the research area of IR, Khurana et al.~\cite{ChatrEx} show that providing NL explanations for why a chatbot did not understand a user's query or does not know what to do next in the context of spreadsheet applications can enhance users' perception of usefulness, transparency, and trust. Additionally, Papenmeier and Topp~\cite{papenmeier2023} show that providing transparency by backchanneling how the users' input was processed by the system can improve their perception of a chatbot's competence as well as their conversational engagement in conversational commerce. Toader et al.~\cite{Toader2019} demonstrate that improving users' perceived competence of a chatbot can enhance their trust in the system.

\subsection{Text Highlighting as an Explanatory Component in Information Retrieval}
\label{subsec:highlighting}
The research field of explainable information retrieval (ExIR) aims to make IR systems more transparent and trustworthy~\cite{Anand}. Text highlighting is used as a simple tool to support users in evaluating the relevance of results in web and scholarly search engines. Searchers rely on descriptions like titles, snippets, and URLs to decide whether they should click and read a particular result~\cite{Zhang2018}. Users’ query terms are highlighted in titles and snippets to give them feedback on where on a website their search terms were found~\cite{Iofciu2009}. The highlighting of text passages as direct answers to user queries is also used in ``feature snippets'', which are integrated into Google's search results~\cite{Strzelecki2020}. Additionally, highlighting can be used to identify relevant passages directly within documents~\cite{Zheng2017}.

Highlighting is also used for local explanations in machine learning. Thus, feature attribution by color highlighting is a common method for explaining the predictions of classification models~\cite{Kim, Schrills}. Recent work investigates how feature attribution can be applied in ranking tasks by highlighting snippets of search queries and document texts~\cite{Anand}.

\subsection{Natural Language Explanations}
\label{subsec:NL}
Natural language (NL) is the most human-like way for systems to provide explanations~\cite{Alonso2020, Cambria2009}. Such explanations are interpretable by users with various backgrounds and levels of technical knowledge~\cite{Mariotti2020}. Juneja et al.~\cite{Juneja2024} conducted a study on search engine explanations with users lacking AI and IR expertise. They found that the participants did not require explanations from a search engine for single, unambiguous answers such as specific dates or values. However, for more complex web search tasks such as product search, the participants deemed concise and easy-to-read NL explanations useful. The authors also note that both text highlighting (in ``feature snippets'') and NL explanations are used by Google's search engine to clarify the search results that were retrieved. Users can access the latter by clicking on the three-dots icon next to a search result in Google. Additionally, Gkatzia et al.~\cite{Gkatzia2016} demonstrate that providing NL explanations for uncertain data such as weather data on a user interface can aid users in decision-making by enhancing their comprehension and informedness, as opposed to solely presenting graphical representations of the data.

\section{Research Question and Hypotheses}
While extensive research has been conducted on explaining the output of RSs, research has not yet established a systematic understanding of the effects of explanations within the result presentation in conversational commerce. In particular, the use of advanced NLP techniques to process user input poses new challenges in terms of transparency. In conversational commerce, a one-to-one mapping of the user's sometimes vague utterances to predefined product attributes is not always obvious compared to what users are used to from traditional faceted search. 

Our research is guided by the following research question:
\begin{questions}
        \item \hypertarget{RQ}{How can explanations that indicate how user utterances were mapped to product attributes support users during the product assessment step in conversational commerce?}
\end{questions}

Inspired by common practices in IR, we investigate two different approaches:

\begin{enumerate}
\item \textbf{Highlighting (H)}: The gathered user preferences are mapped to product attributes, which are then highlighted in the result presentation (see Fig.~\ref{fig:f1}).
\item \textbf{Highlighting + Explanation (H+E)}: In addition to the highlighting, the mapping is explained in natural language (see Fig.~\ref{fig:f2}).
\end{enumerate}

To answer our research question, we conducted an online user study using a between-subjects design with three conditions (H, H+E, B~= Baseline without any explanations). To limit complexity and inspired by Google’s use of highlighting in conjunction with NL explanations (see Section~\ref{subsec:NL}), we decided not to include a condition in which only NL explanations are provided. We assume that explanations positively affect perceived system transparency, usefulness, and users' confidence in their purchase decisions. Table~\ref{tab:Hypotheses} presents our hypotheses and the underlying assumptions.

\begin{table*}[]
\centering
\caption{Research hypotheses and underlying assumptions. Dependent variables are formatted in bold.}
\label{tab:Hypotheses}
\small
\begin{tabular}{p{0.65\textwidth} p{0.35\textwidth}}
\toprule[1.2pt]
\textbf{Assumption} & \textbf{Hypothesis} \\
\midrule
Jin et al. \cite{Jin2024} show that explanations provided within the chat positively influence users' perceived transparency in conversational commerce. Similarly, in traditional RSs, literature shows that explanations also enhance perceived transparency~\cite{Cramer2008,Gedikli,Framework,Vig} (see Section~\ref{subsec:effects}). We assume that those findings likewise hold true for the result presentation in conversational commerce. & \begin{H1}\label{hyp:1}Users who are provided with H+E have a higher \textbf{perception of transparency} than those with H only. Users assigned to B have the lowest perception of transparency.\end{H1} \\
\midrule
Vig et al.~\cite{Vig} and Verbert et al.~\cite{Verbert} show that explanations can increase decision effectiveness in RSs by allowing users to make more informed decisions. We assume that enabling users to make more informed decisions also increases their decision-making confidence after assessing a recommended item in conversational commerce. As highlighting can be categorized as a type of explanation~\cite{Juneja2024}, albeit more rudimentary than NL explanations, we assume that it leads to a higher confidence than the baseline condition. & \begin{H1}\label{hyp:2}Users who are provided with H+E have the highest \textbf{confidence} in their decision. The lowest confidence is reported by users in B.\end{H1} \\
\midrule
Pu et al. \cite{Framework} show that by improving perceived transparency, confidence, and trust, explanations can increase users' purchase intentions in RSs. Furthermore, Zhou~\cite{Zhou2015} found that by increasing customers' trust, perceived transparency can improve their purchase intentions in online stores. We assume that these findings can be transferred to or are applicable to the result presentation in conversational commerce. & \begin{H1}\label{hyp:4}Users who are provided with H+E have a higher \textbf{purchase intention} compared to both H and B.\end{H1} \\
\midrule
Explanations were shown to increase decision-making effectiveness in RSs~\cite{Vig, Verbert}. We assume that these findings can be transferred to the result presentation in conversational commerce. However, while effectiveness is specifically related to decision-making (as also described by Tintarev \cite{Tintarev2007}), see Section~\ref{subsec:goals}), we use the related variable ``perceived usefulness'' to examine the users' perception of the assistant as a whole. We posit that enhanced decision-making effectiveness also leads to an increase in the perceived usefulness of the assistant. This aligns with research by Ing and Ming \cite{Ing2018} who use shopping effectiveness as one of several items to measure the perceived usefulness of blogger recommendations. & \begin{H1}\label{hyp:5}Users who are provided with H+E perceive the digital advisor as most \textbf{useful}. B is perceived the least useful.\end{H1} \\
\midrule
Khurana et al.~\cite{ChatrEx} show that NL explanations of errors provided by a chatbot in a spreadsheet tool enhance users' perception of usefulness. We assume that this finding can be applied to the result presentation in conversational commerce. & \begin{H1}\label{hyp:6}H+E is perceived as more \textbf{useful} than H.\end{H1} \\
\bottomrule[1.2pt]
\end{tabular}
\end{table*}

\section{Study Design}
\subsection{Apparatus - Digital Product Advisor}
\begin{figure*}
    \centering
    \includegraphics[width = 0.75\textwidth]{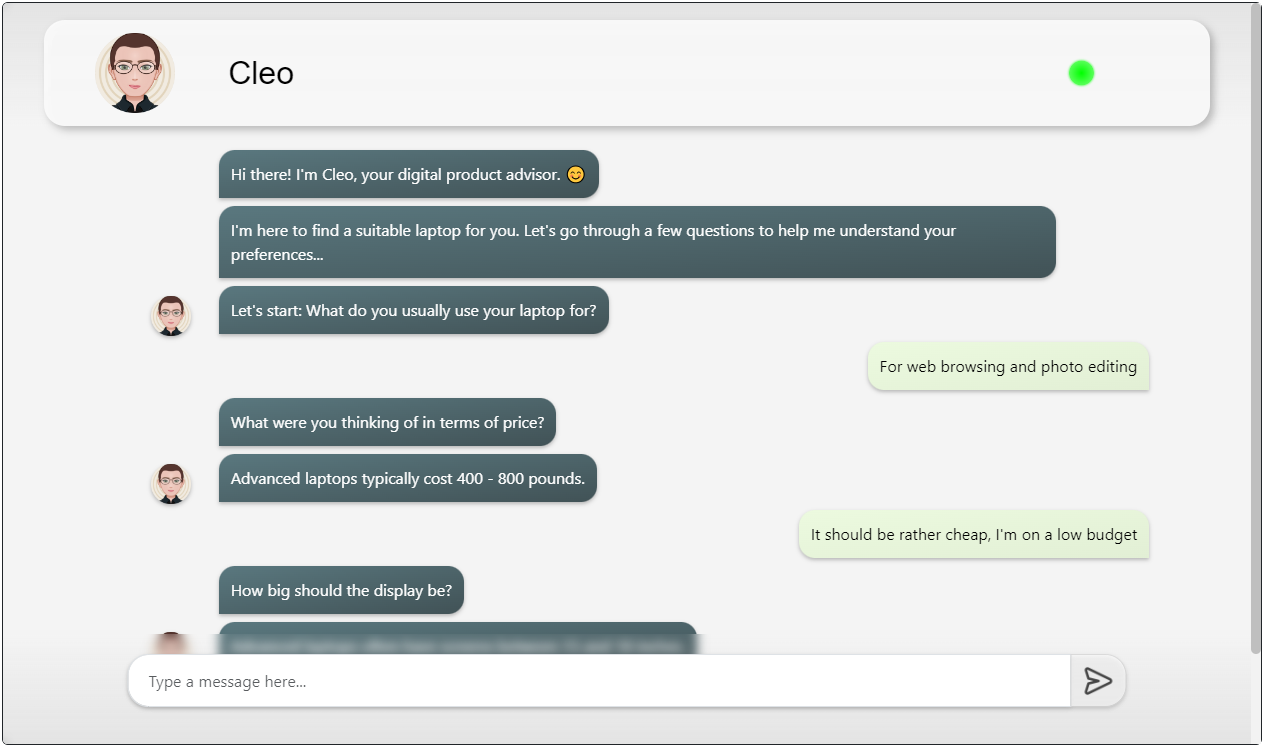}
    \caption{Interface for our digital assistant}
    \label{fig:Interface}
    \Description{Screenshot of a chat message exchange on a computer.}
\end{figure*}
We developed a rule-based digital assistant for laptop search, as shown in Fig.~\ref{fig:Interface}. The underlying dataset was collected from Amazon\footnote{\url{https://www.amazon.com/}} and contains 3,638 laptop descriptions. The digital assistant queries users about their preferences for different laptop aspects in a predefined order that aims to mimic the questioning behavior of a human shop assistant: Purpose, price, screen size, hard drive storage, RAM size, and battery life. The user's response regarding the intended purpose gets mapped to either basic, advanced, or gaming and is used to provide recommendations for the subsequent attributes. After each attribute, the system continuously filters the dataset for suitable laptops by mapping the user's responses to specific values. Research has shown that the vagueness of user queries in the context of product search varies to a high degree~\cite{Papenmeier2020}. Therefore, our assistant uses a BERT model and an LSTM model to identify vague user statements (e.g., ``rather small'' or ``not too big'') and map them to specific value ranges (e.g., 12 to 14 inches) for the respective product attributes. Especially when vague user input is provided, we expect explanations to help users understand why a particular product is recommended.

After inquiring about all the described laptop aspects, our system presents a suitable laptop to the user. If multiple laptops in the dataset match the user's preferences, the one with the highest average user rating is suggested. However, if only one fitting laptop remains earlier in the conversation, the assistant ends the conversation and moves on to the result presentation. To simulate a first suggestion from a human shop assistant, we have chosen to display only one result. Our setup focuses on providing explanations for this initial suggestion. The suitable laptop is presented as a typical online shop result page with a title, user rating, product image, and product description.

\begin{figure*}
  \centering
  \subfloat[Highlighting (H)]{
    \includegraphics[width=0.75\textwidth]{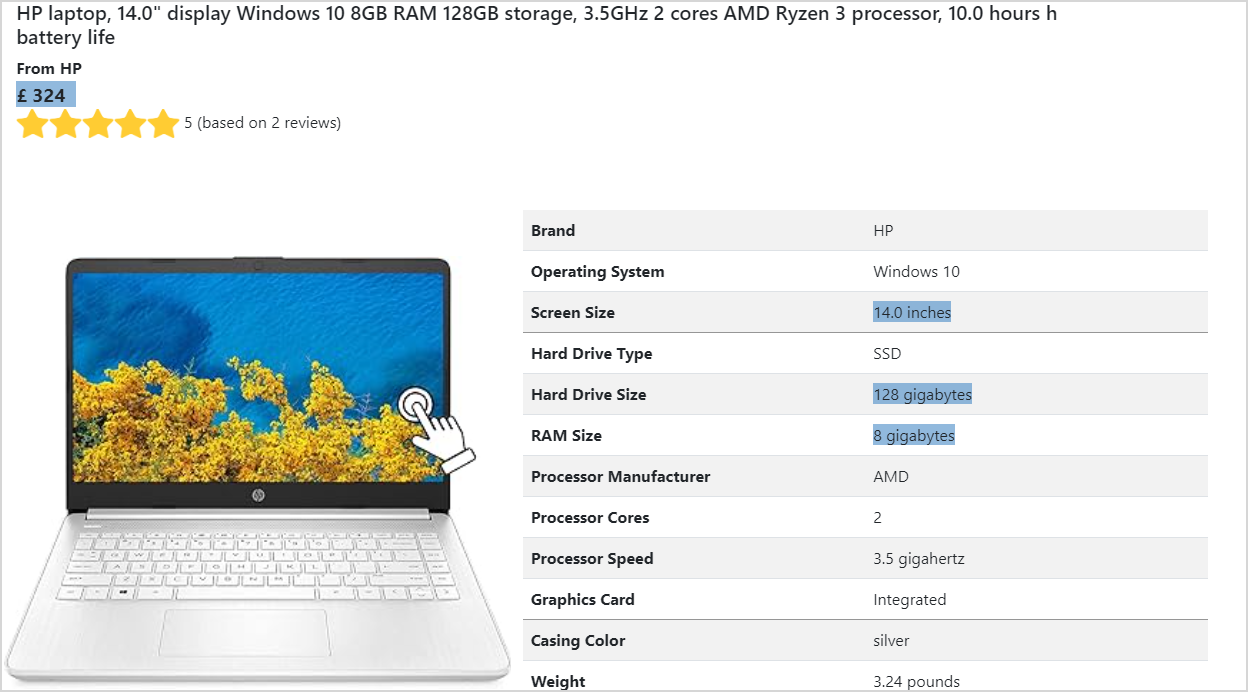}
    \label{fig:f1}
    \Description{Screenshot of a product presentation that includes a laptop image and product details in the form of technical information displayed above and next to it. Text snippets of the product details that match the user's preferences are highlighted in color.}
  }

  \subfloat[Highlighting plus natural language explanations (H+E)]{
    \includegraphics[width=0.75\textwidth]{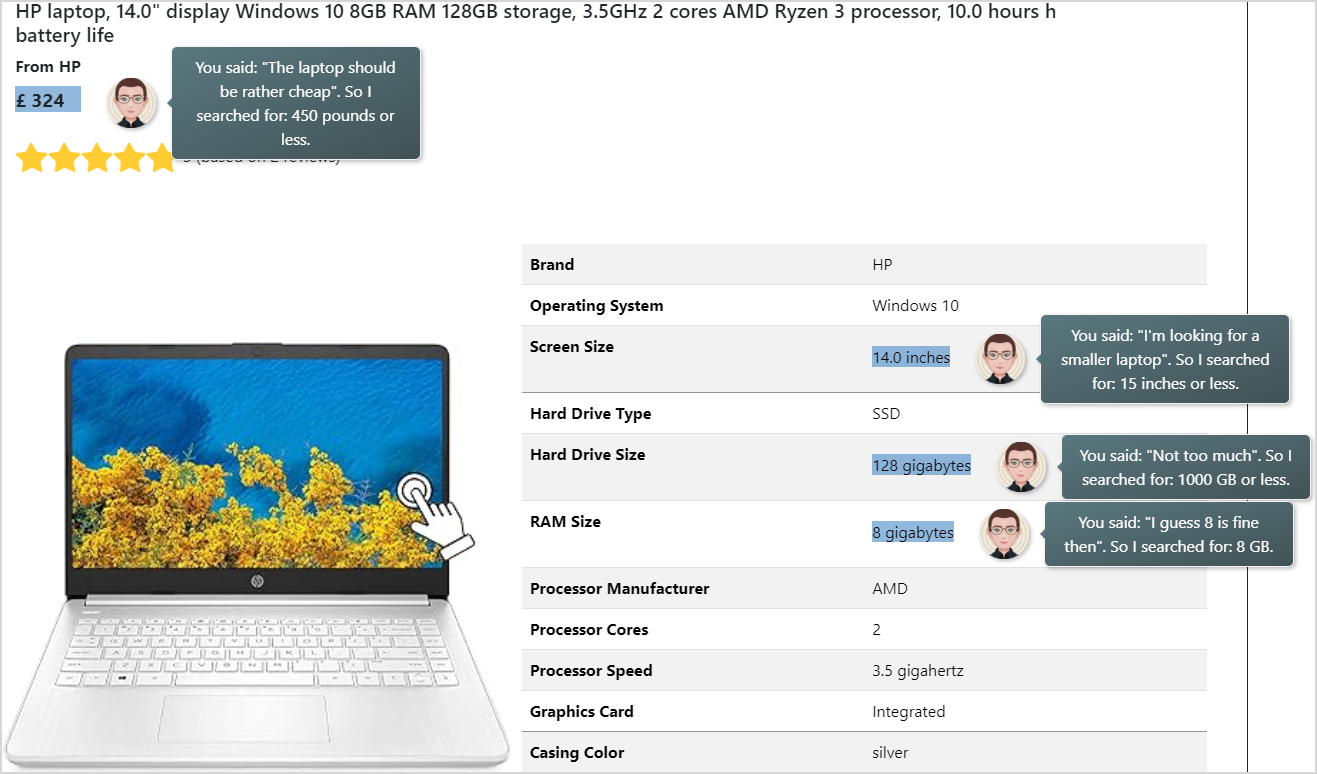}
    \label{fig:f2}
    \Description{Screenshot of a product presentation that includes a laptop image and product details in the form of technical information displayed above and next to it. Text snippets of the product details that match the user's preferences are highlighted in color. In addition, speech bubbles are displayed next to the attributes, citing the user's stated preference and describing how it influenced the digital assistants' search.}
  }
  \caption{Explanation variants}
  \label{fig:Variants}
\end{figure*}

We tested two different explanation variants to present the results showing how a user's utterances relate to different aspects of the recommended laptop alongside a baseline condition (B) without any explanation. In the first variant (H), the aspects the user was asked about are highlighted in color (see Fig.~\ref{fig:f1}). Given the common usage of highlighting (see Section~\ref{subsec:highlighting}), we adopted it as a straightforward tool in our user study to investigate its effects on user perception in the context of conversations commerce. In the second variant (H+E), our assistant offers NL explanations in the form of preference repetition to the user in speech bubbles (see Fig.~\ref{fig:f2}). The explanation consists of a citation of the user's response regarding a laptop aspect they were asked about (``You said: [...].'') and a description of how it impacted the assistant's search (``So I searched for [...].''). An example would be ``You said: `The laptop should be rather small and portable as I move around a lot at work'. So I searched for: 15 inches or less.'' Inspired by faceted search and search engine user interfaces, we deliberately integrate several aspect-specific explanations into the result presentation instead of providing a single, summative explanation within the chat (like in Jin et al.'s~\cite{Jin2024} ``PhoneBot'', see Section~\ref{subsec:effects}). This way, users can find the explanations exactly next to the product attributes they are referring to.

\subsection{Procedure and Scenario}
We set up an online study on SoSci Survey\footnote{\url{https://www.soscisurvey.de/en/index}}. Participants first provide consent and answer closed questions about their demographics. They then rate their domain knowledge for laptops on a scale from 1 (no knowledge/non-expert) to 7 (high knowledge/expert) before being presented with the scenario and task description (adapted from~\cite{Papenmeier2021}): \\

\say{\textit{Imagine that your laptop stopped working, and you are now searching for a new one.
The advisor will inquire about your preferences and requirements to suggest a suitable laptop for you. Your task will be to interact with the advisor, examine the recommended laptop and decide whether you would buy it.}} \\

Participants are then randomly assigned and directed to one of the three interface variants (B, H, H+E), receive a reminder of the scenario, and start interacting with the digital assistant (see Fig. \ref{fig:Interface}) until they receive their laptop recommendation (as shown in Fig. \ref{fig:Variants} according to their condition). Upon scrolling down to view the complete product overview that includes all product aspects that the participants were asked about, a button labeled ``I have made a decision regarding the purchase of the laptop. Take me to the next step!'' appears on the interface. After clicking this button, participants are asked about the likeliness of buying the recommended laptop and subsequently about their confidence in their previous answer. In case of low likeliness to buy, they are asked for additional feedback with an open question. Thereafter, participants are redirected to our questionnaire and answer open and closed questions about their experience with the digital assistant. Our institute's ethics committee has confirmed the study's ethical clearance.

\subsection{Measurements}
This section presents the dependent variables that were tested and the types of data that were gathered. Tables provide insight into how we measured the dependent variables.

\subsubsection{Measures for Perceived Transparency, Confidence, and Purchase Intention}
\begin{table*}[]
\centering
\caption{Dependent variables measured with 5-point Likert scale questions or statements}
\label{DependentVars1}
\begin{tabular}{p{0.2\textwidth} p{0.54\textwidth} p{0.21\textwidth}}
\toprule[1.2pt]
\textbf{Dependent variable} & \textbf{Question/Statement} & \textbf{Likert Scale Options} \\
\midrule
Perceived transparency & {\raggedright The product advisor explained why the product was \\ recommended to me. (\textbf{\hypertarget{PT1}{PT1}})} & {\raggedright Strongly disagree (1) -- \\ Strongly agree (5)} \\
\cmidrule{2-3}
 & I understood why the product was recommended to me. (\textbf{\hypertarget{PT2}{PT2}}) & {\raggedright Strongly disagree (1) -- \\ Strongly agree (5)} \\
\cmidrule{2-3}
 & {\raggedright The product advisor provided information about how my \\ preferences were considered.} (\textbf{\hypertarget{PT3}{PT3}}) & {\raggedright Strongly disagree (1) -- \\ Strongly agree (5)} \\
\cmidrule{2-3}
 & How easy was it for you to understand why the product was recommended to you? (\textbf{\hypertarget{PT4}{PT4}}) & {\raggedright Very difficult (1) -- \\ Very~easy~(5)} \\
\midrule
Confidence & How confident are you in your decision? & {\raggedright Not at all confident (1) -- \\ Extremely confident (5)} \\
\midrule
Purchase intention & How likely is it that you would buy this laptop? & {\raggedright Extremely unlikely (1) -- \\ Extremely likely (5)} \\
\bottomrule[1.2pt]
\end{tabular}
\end{table*}

We took measures for the dependent variables confidence and perceived transparency. Additionally, we measured purchase intention, i.e., the users' likeliness to buy the recommended laptop. All mentioned variables were measured with 5-point Likert scale questions or statements (cp. Table~\ref{DependentVars1}). The statements for \hyperlink{PT1}{PT1}, and \hyperlink{PT2}{PT2} are inspired by previous work from Jin et al.~\cite{Jin2024}, while the one for \hyperlink{PT3}{PT3} is inspired by Hellman et al.~\cite{Hellmann}. Peters~\cite{Peters2022} inspired the statement for confidence.

\subsubsection{Measures for Usefulness}
We also measured users' perceived usefulness of the digital assistant for all conditions (statement inspired by Pu et al.~\cite{Framework}) as well as the usefulness for H and H+E in particular. For the interface variants H and H+E, we asked additional specific questions that were not applicable for the baseline condition. We took measures for the perceived usefulness of the highlighting and the explanations, respectively. For H+E, we also measured the explanations' understandability (cp. Table~\ref{DependentVars2}). We again used a five-point Likert scale with statement and questions.

\begin{table*}[]
\centering
\caption{5-point Likert scale statements or questions for the interface variants H and H+E}
\label{DependentVars2}
\begin{tabular}{p{0.2\textwidth} p{0.54\textwidth} p{0.21\textwidth}}
\toprule[1.2pt]
\textbf{Dependent variable} & \textbf{Question/Statement} & \textbf{Likert Scale Options} \\
\midrule
{\raggedright Perceived usefulness} & {\raggedright The advisor helped me find a suitable product.} & {\raggedright Strongly disagree (1) -- \\ Strongly agree (5)} \\
\midrule
{\raggedright Perceived usefulness (H, H+E)} & {\raggedright How useful did you find the [highlighting/explanations] while forming an opinion on the recommended product?} & {\raggedright Not at all useful (1) -- \\ Extremely useful (5)} \\
\midrule
{\raggedright Understandability (only for H+E)} & {\raggedright How easy was it for you to understand the language of the \\ explanations?} & {\raggedright Very difficult (1) -- \\ Very~easy~(5)} \\
\bottomrule[1.2pt]
\end{tabular}
\end{table*}

\subsubsection{Qualitative User Feedback}
As the quantitative variables do not give us detailed information about why people like some things and dislike others, we included some open-ended questions to gather more individual feedback. This also helps us to identify areas for improvement for future work. All open questions are listed in Table~\ref{tab:Qualitative}. \hyperlink{Q5}{Q5} was asked after the perceived usefulness rating for H and H+E (cp. Table~\ref{DependentVars2}).
\begin{table}[]
    \centering
    \caption{Open-ended questions for eliciting qualitative data}
    \label{tab:Qualitative}
    \begin{tabular}{p{\columnwidth}}
    \toprule[1.2pt]
    \textbf{Open-ended questions} \\
    \midrule
        \hypertarget{Q1}{What specifically confused you about the recommendation?} (\textbf{Q1}) \\
        \hypertarget{Q2}{Which specific aspect(s) from your interaction with the product advisor stood out to you?} (\textbf{Q2}) \\
        \hypertarget{Q3}{What (other) explanations would you like to be provided by the advisor?} (\textbf{Q3}) \\
        \hypertarget{Q4}{What specifically helped you understand the recommendation?} (\textbf{Q4}) \\
        \hypertarget{Q5}{Please justify your rating [for highlighting/explanation usefulness] above.} (\textbf{Q5}) \\
    \bottomrule[1.2pt]
    \end{tabular}
\end{table}

\subsection{Participants}
We recruited 135 participants (68 male, 66 female, 1 inter/non-binary) through the online crowdsourcing platform Prolific\footnote{\url{https://www.prolific.com/}}. During the course of the study, 14 participants either chose to withdraw or had their participation automatically canceled by Prolific after 56 minutes. In these cases, Prolific automatically recruited new participants until a total of 135 study completions was reached. They had a mean domain knowledge of 4.4, with a standard deviation of 1.5. A Kruskal-Wallis test showed that there was no significant difference (p~=~0.3584) in domain knowledge between conditions. The participants (ages 18-91, M~=~40.9, STD~=~13.5) were randomly assigned to one of the three conditions (B~=~47, H~=~43, H+E~=~45). To be eligible, they had to be located in the UK, speak English as their primary language, and not have any language difficulties. Participation in the study took approximately 11 minutes per participant (M~=~10.53~min, STD~=~5.22~min). Each participant who completed the study received a compensation of 2.25~GBP, corresponding to an hourly rate of 12.27~GBP.

\section{Results}
This section presents our findings and the data collected during our user study. First, we describe the results of our statistical analyses and evaluate our hypotheses. We then present the qualitative insights we gathered from the participants' answers to the open-ended questions.

\subsection{Statistical Analysis}
To test for significant differences between conditions, we used the Kruskal-Wallis test with a Dunn's test as post-hoc analysis, applying the Bonferroni correction to account for the multiple testing bias. To compare only two conditions, we used the Mann-Whitney U test. We used non-parametric tests because the survey data was not normally distributed. For analyzing correlations, we used the measure of Spearman’s rho. Means and standard deviations for our dependent variables are presented in Table~\ref{tab:Values}.

\begin{table*}[h]
\centering
\caption{Mean and standard deviation for each dependent variable. Variables for which significant differences were found are formatted in bold.}
\label{tab:Values}
\begin{tabular}{|l|r|r|r|r|r|r|}
\hline
 & \multicolumn{2}{c|}{\textbf{Baseline}} & \multicolumn{2}{c|}{\textbf{Highlighting}} & \multicolumn{2}{c|}{\textbf{H+E}} \\
\cline{2-7}
 & \multicolumn{1}{c|}{Mean} & \multicolumn{1}{c|}{SD} & \multicolumn{1}{c|}{Mean} & \multicolumn{1}{c|}{SD} & \multicolumn{1}{c|}{Mean} & \multicolumn{1}{c|}{SD} \\
\hline
\textbf{PT1} & \textbf{3.04} & \textbf{1.00} & \textbf{3.26} & \textbf{1.20} & \textbf{3.96} & \textbf{0.77} \\
PT2 & 4.00 & 0.81 & 4.09 & 1.09 & 4.29 & 0.63 \\
\textbf{PT3} & \textbf{3.30} & \textbf{1.08} & \textbf{3.42} & \textbf{1.22} & \textbf{4.13} & \textbf{0.76} \\
\textbf{PT4} & \textbf{3.96} & \textbf{0.69} & \textbf{4.00} & \textbf{0.90} & \textbf{4.38} & \textbf{0.72} \\
Confidence & 3.79 & 0.86 & 4.02 & 0.80 & 4.00 & 0.90 \\
Purchase intention & 3.19 & 1.10 & 3.23 & 1.41 & 3.36 & 1.23 \\
\hline
Perceived usefulness & 3.70 & 0.98 & 3.60 & 1.28 & 3.89 & 0.91 \\
Perceived usefulness (H, H+E) & - & - & 3.70 & 1.24 & 3.67 & 1.02 \\
Understandability (H+E) & - & - & - & - & 4.62 & 0.53 \\
\hline
\end{tabular}
\end{table*}

\begin{table*}[h]
\centering
\caption{Results of the Kruskal-Wallis tests comparing B, H, and H+E. Significant results are formatted in bold.}
\label{KruskalWallisResults}
\begin{tabular}{p{0.18\textwidth} r r r}
\toprule[1.2pt]
\textbf{Dependent~variable} & \textbf{p-value (adjusted)} & \textbf{Test Statistic (H)} & \textbf{Effect Size (η2)} \\
\midrule
\textbf{PT1} &  \textbf{0.00009} & \textbf{18.6911} & \textbf{0.1300} \\
PT2 & 0.1518 & 3.7708 & 0.0130 \\
\textbf{PT3} & \textbf{0.0004} & \textbf{15.7999} & \textbf{0.1000} \\
\textbf{PT4} & \textbf{0.0091} & \textbf{9.3828} & \textbf{0.0560} \\
Confidence & 0.3474 & 2.1148 & 0.0009 \\
Purchase intention & 0.7729 & 0.5152 & -0.0110 \\
\midrule
Perceived usefulness & 0.6294 & 0.926 & -0.0081 \\
\bottomrule[1.2pt]
\end{tabular}
\end{table*}

\subsubsection{Effects of Explanations on Perceived Transparency, Confidence, and Purchase Intention}
\begin{figure*}
  \centering
  \subfloat[PT1]{
    \includegraphics[width=0.32\textwidth]{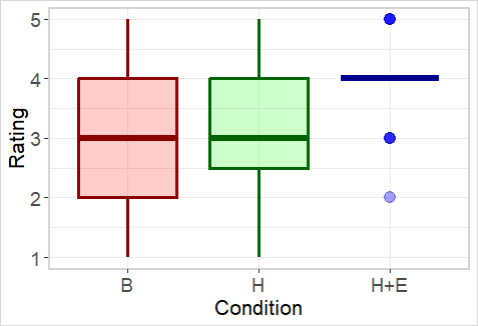}
    \label{PT1}
    \Description{Boxplots visualizing the distribution of user responses for the variable PT1 for each of the three conditions. Mean and standard deviation values are provided in Table 5, column 1. H+E has the highest mean value, while B and H have lower and similar mean values. H+E has some outliers.}
  }
\hfill
  \subfloat[PT3]{
    \includegraphics[width=0.32\textwidth]{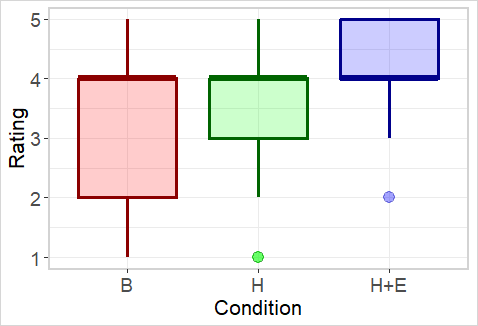}
    \label{PT3}
    \Description{Boxplots visualizing the distribution of user responses for the variable PT3 for each of the three conditions. Mean and standard deviation values are provided in Table 5, column 3. While all conditions have similar mean values, the interquartile range of H+E is higher than both B and H.}
  }
  \hfill
  \subfloat[PT4]{
    \includegraphics[width=0.32\textwidth]{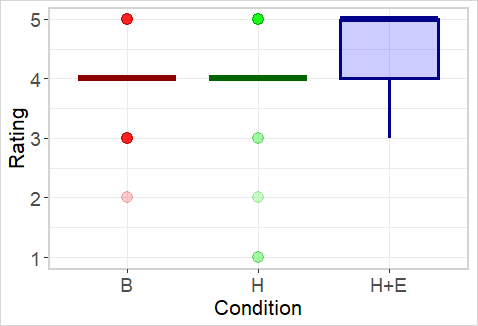}
    \label{PT4}
    \Description{Boxplots visualizing the distribution of user responses for the variable PT4 for each of the three conditions. Mean and standard deviation values are provided in Table 5, column 4. H+E has the highest mean value, while B and H have lower and similar mean values. Both B and H have some outliers.}
  }  
  \caption{Significant Kruskal-Wallis test results for PT1, PT3, and PT4}
\end{figure*}

The Kruskal-Wallis tests showed no significant differences (p~>~0.05) between conditions for \hyperlink{PT2}{PT2} (\textbf{H\ref{hyp:1}} rejected for \hyperlink{PT2}{PT2}), confidence (\textbf{H\ref{hyp:2}} rejected), purchase intention (\textbf{H\ref{hyp:4}} rejected), and perceived usefulness (\textbf{H\ref{hyp:5}} rejected) (cp. Table~\ref{KruskalWallisResults}). Significant differences between groups were found for \hyperlink{PT1}{PT1}, \hyperlink{PT3}{PT3}, and \hyperlink{PT4}{PT4}. Post-hoc Dunn's tests were conducted to determine which specific conditions differed significantly. For \hyperlink{PT1}{PT1} (see Fig.~\ref{PT1}), H+E was rated significantly higher than both B (p~<~0.001) and H (p~<~0.01). For \hyperlink{PT3}{PT3} (see Fig.~\ref{PT3}), H+E was also rated significantly higher than B (p~<~0.001) and H (p~<~0.01). Finally, for \hyperlink{PT4}{PT4} (see Fig. \ref{PT4}), H+E received a significantly higher score than B (p~<~0.01). However, no significant difference was found between H+E and H (p~=~0.0527). Thus, \textbf{H\ref{hyp:1}} can be partially accepted for \hyperlink{PT1}{PT1}, \hyperlink{PT2}{PT2}, and \hyperlink{PT4}{PT4}.

\subsubsection{Perceived Usefulness and Understandability}
A Kruskal-Wallis test revealed no significant differences (p~>~0.05) between conditions for perceived usefulness (cp. Table~\ref{KruskalWallisResults}). A Mann-Whitney U test further showed no significant difference in perceived usefulness (H, H+E) between H and H+E (W~=~1033.5, p~=~0.5621). Thus, \textbf{H\ref{hyp:6}} has to be rejected. For explanation understandability, the mean was 4.62 and the standard deviation 0.53 (cp. Table~\ref{tab:Values}).

\subsubsection{Effects of Domain Knowledge on Confidence}
As reported above, we did not find a significant main effect of decision-making confidence between conditions. However, when categorizing all participants in two groups of below-average (1-3) and above-average (5-7) domain knowledge, a Mann-Whitney U test showed that the confidence of participants with above-average domain knowledge was significantly higher (p~<~0.01) than of those with below-average domain knowledge. We also found a weak positive correlation (r~=~0.3024) between domain knowledge and confidence.

\subsection{Qualitative User Feedback}
\label{subsec:qualitative}
We analyzed participants' responses to the open-ended questions (cp. Table~\ref{tab:Qualitative}) through thematic analysis \cite{Clarke2017}, using codes to inductively identify patterns and common themes. Regarding participants' responses to \hyperlink{Q1}{Q1}, it seems that the laptop recommendation is perceived as slightly less confusing in H and H+E. 79\% (H) and 71\% (H+E) of the participants answered this question with ‘nothing,’ in contrast to 53\% in the baseline condition. The other statements in all conditions, among the single statements, are about complaining about incorrect processing of user input (6\%) and the lack of alternative laptop choices (7\%). In the baseline group, 4\% of the participants explicitly asked for an explanation of why the recommended item was suggested. 

Feedback on what stood out to participants (\hyperlink{Q2}{Q2}) can be divided into feedback on the conversation and feedback on the result or the presentation of the result. The speed of the conversation was praised by 12\% of all participants; 10\% liked the suggestions and explanations given by the digital advisor during the conversation. 4\% liked the conversation in general, while 2\% complained about not being able to ask questions, and 4\% would have liked to be asked more questions. Regarding the results and presentation of the results, again, 3\% complained about the lack of choice. In both the H and the H+E conditions, 13\% of participants explicitly stated that they liked the recommended item (e.g., participant P103 wrote ``that [the advisor was] able to find a product that fit all my needs''). In the baseline condition, zero participants left an explicit comment about this when asked what stood out to them. 51\% of the participants in the baseline condition stated that the layout and the presentation itself helped them to understand the recommended item \hyperlink{Q4}{Q4}. In the H and H+E conditions, this was mentioned less (H: 30\%, H+E: 11\%). They were more likely to find the highlighting (H: 28\%, H+E: 9\%) and the explanations helpful (H+E: 64\%). P22, for example, explained that they liked ``The pop up boxes explaining why the product had been chosen for me''.

When asked for (further) explanations (\hyperlink{Q3}{Q3}), B:~25\%, H:~42\% and H+E:~27\% answered ``none''. In the B and H conditions, 23\% and 26\%, respectively, explicitly asked for an explanation as to why the recommended item was chosen (e.g., P23: ``Would like to understand why this specific product was chosen [...]''). 11\% of the participants in the H+E condition asked for more choices and explanations of their differences. Justifying why a result was chosen contributed to perceived usefulness (\hyperlink{Q5}{Q5}) for 22\% of participants in the H+E condition (e.g., P140: ``It justifies why the product was selected for me. It allows me to see how and why the product may or may not be suitable for my needs''). On the other hand, 15\% of the participants stated that the NL explanations described the obvious. Highlighting was mentioned as useful by 40\% in the H condition. 21\% liked it because it drew attention (e.g., P125: ``Helps to see if it met my requirements''). Highlighting was not explicitly mentioned at all in the H+E condition.

Across conditions, 30\% of participants responded that it is either unlikely or extremely unlikely that they would buy the laptop recommended to them. 37\% of them noted that they would have preferred a different operating system and/or laptop brand, while 27\% mentioned other product specifications would prevent them from buying (e.g., P141: ``Because of the graphics card''). Additionally, 24\% stated that the digital assistant did not understand their preferences correctly (e.g., P23: ``Because I asked for a 17 inch laptop, and the screen size is 16 inches''). Two participants (5\%) mentioned that the laptop's aesthetics did not appeal to them, while P10 stated that they ``feel as if the interaction was not real enough and [doesn't] give [them] the most confidence in choosing a laptop to buy''. B:~21\%, H:~13\%, and H+E:~8\% mentioned a lack of product information displayed on the interface.

\section{Discussion}
\label{discussion}
The results indicate that the NL explanations provided by our digital assistant can support users during their assessment of a recommended product by significantly improving their perception of transparency. Our findings are in line with those from the RS domain \cite{Framework, Cramer2008, Gedikli, Millecamp, Vig}, which show that explanations can enhance transparency as perceived by users (see Section~\ref{subsec:effects}). 

According to qualitative feedback, users found the NL explanations to be particularly helpful for understanding why an item was recommended (see Section~\ref{subsec:qualitative}). This suggests that the explanations improved the effectiveness of the recommendation, which Tintarev and Masthoff~\cite{Tintarev2007} defined as one goal of explanations in RSs (see Section~\ref{subsec:goals}). Additionally, participants' feedback indicates that reducing their confusion about the recommendation was one factor that enhanced their perception of transparency. We assume that the personalized explanations improved the transparency factors ``User Representation'' and ``System Parameters and Logic'' defined by Vorm and Miller~\cite{Vorm2018} (see Section~\ref{subsec:goals}). Meanwhile, highlighting did not significantly affect any of the dependent variables. Qualitative feedback indicates that it did not assist participants in deducing why a recommended item was selected. 

Although participants from H+E reported finding the NL explanations helpful in the open-ended questions, the quantitative findings show that they do not improve users' confidence in their decision-making. Thus, our study on conversational commerce could not transfer Pu et al.'s~\cite{Framework} finding of a significantly positive effect of perceived transparency on users’ confidence in RSs to the result presentation in conversational commerce. Qualitative user feedback suggests that users' confidence was influenced by several confounding factors, including reliance on or preference for a certain operating system or brand, the inability to compare results due to the suggestion of only one laptop, unsatisfactory or missing product information, and qualitative aspects such as aesthetic preferences. The desire of participants to compare multiple items can be attributed to the ``Options'' factor in Vorm and Miller's~\cite{Vorm2018} five-factor model of transparency in RSs (see Section~\ref{subsec:goals}). Analysis of user input showed that participants who mentioned that the NL explanations stated the obvious mostly responded with concrete values rather than vague statements. This suggests that our product advisor was particularly helpful to those participants who did not have a precise idea of the laptop they wanted. Conversely, this indicates that it may be sufficient to show explanations for user utterances in which vagueness was detected.

While the presence of NL explanations did not significantly affect users’ perceived understanding of why a product was recommended (\hyperlink{PT2}{PT2}), it did significantly improve the perceived ease of understanding (\hyperlink{PT4}{PT4}). This suggests that while NL explanations may not enhance users’ confidence in their understanding of the recommendation, they do make the process of understanding why a product was recommended easier. We assume that the explanations may serve to reduce the cognitive effort required to understand recommendations.

While some participants in H and H+E explicitly mentioned in the open questions that they liked the recommended item, none mentioned this in the baseline condition. This would be consistent with the persuasive effects of explanations, as described by Masthoff and Tintarev~\cite{Tintarev2007} (see Section~\ref{subsec:goals}). However, we could not underscore this finding through a statistically significant difference in participants' purchase intention. We assume this to be a consequence of the explanations' focus on system transparency rather than result justification (see Section~\ref{relatedWork}). Furthermore, we assume that the usability limitations of our digital assistant, such as the lack of choices and users' inability to ask questions or adjust their preferences, had a halo effect on the participants' emotional responses~\cite{Minge2018}, which may have confounded their assessment of both decision-making confidence and purchase intention. We will address this issue in future work by improving the digital assistant based on our participants' feedback.

\section{Conclusion and Future Work}
With this study, we explored how explanations can support customers during product assessment in the context of conversational commerce. Focusing on the laptop domain, we investigated users' perceptions when provided with text snippet highlighting and NL in addition to this highlighting as explanation variants for recommended products. The explanations aimed at compensating for the absence of facets and account for the use of advanced NLP techniques in conversational commerce. Our study suggests that providing NL explanations and text snippet highlighting for a recommended product supports users during item assessment by enhancing their perception of transparency. Furthermore, NL explanations were deemed helpful by users. On the other hand, highlighting without NL explanations could not support users. These findings can assist online retailers in promoting transparency during the product evaluation step when developing user interfaces for this growing field of e-commerce.

In future work, we plan to investigate how users can be further supported in their result assessment within conversational commerce. As many participants of our study wished to be presented with several results to compare rather than just one, we will expand our user interface by adding more results and exploring how to display result lists and explain differences between items to further assist users in evaluating products. Additionally, we will consider our findings that users with limited domain knowledge have less decision-making confidence than those with above-average domain knowledge. We will investigate how to personalize conversational interactions to provide appropriate support to users based on their domain knowledge. As our study had limitations in this regard, users' decision-making confidence could be retested with a more advanced digital assistant that enables them to state their preferences for additional product aspects. Furthermore, as some participants complained about incorrect input processing, a logical next step is to leverage the scrutiny goal of explanations by allowing users to continue the conversation beyond the initial result presentation. This would enable them to clarify any misunderstandings and adjust their preferences. Lastly, we did not examine the isolated effects of NL explanations on user perception, as this was not the focus of our study. Future research could investigate the independent effects of explanation formats (highlighting vs. standalone NL) to gain a more comprehensive understanding.

\begin{acks}
This work was supported by the German Research Foundation (DFG) as part of the "VACOS 2" project (no. 388815326). We thank our study participants and our anonymous reviewers for their helpful feedback. We would also like to thank Niklas Kerkfeld for his assistance with the technical implementation.
\end{acks}

%References

%% the liking and willingness of users to download or buy a recommended item \cite{Medhurst} and to significantly influence users' trust and confidence \cite{Framework} in RSs.

%% BIBLIOGRAPHY
\bibliographystyle{ACM-Reference-Format}
\bibliography{cui24-bubbles-bib}

\end{document}